\documentclass[onecolumn,pre,a4paper]{revtex4}

\usepackage{graphicx}
\usepackage{amsmath}
\usepackage{latexsym}
\usepackage{psfrag}

\newcommand{\bq}{\begin{equation}}
\newcommand{\eq}{\end{equation}}
\newcommand{\ba}{\begin{eqnarray}}
\newcommand{\ea}{\end{eqnarray}}
\newcommand{\p}{\mathbf{p}}
\newcommand{\pp}{\mathbf{P}}
\newcommand{\A}{\mathbf{A}}

\newcommand{\tr}{\tau}

\begin{document}
\title{\bf Homogeneous complex networks}
\author{Leszek Bogacz}\thanks{Email address: bogacz@th.if.uj.edu.pl}
\author{Zdzis\l{}aw Burda}\thanks{Email address: burda@th.if.uj.edu.pl}
\author{Bart\l{}omiej Wac\l{}aw}\thanks{Email address: 
bwaclaw@th.if.uj.edu.pl}

\affiliation{Mark Kac Center for Complex Systems Research and
Marian Smoluchowski Institute of Physics, \\
Jagellonian University, Reymonta 4, 30-059 Krak\'ow, Poland, \\
Institut f\"ur Theoretische Physik, 
Universit\"at Leipzig, Augustusplatz 10/11, D-04109 Leipzig, Germany}

\begin{abstract}
We discuss various ensembles of homogeneous complex
networks and a Monte-Carlo method of generating graphs from 
these ensembles.
The method is quite general and can be applied to simulate 
micro-canonical, canonical or grand-canonical ensembles 
for systems with various statistical weights. 
It can be used to construct
homogeneous networks with desired properties, or to construct 
a non-trivial scoring function for problems of advanced motif 
searching.
\end{abstract}

\maketitle

\section{Introduction}

Complex networks is a new emerging branch of random graph theory. 
For a long time random graphs have been mainly
studied by pure mathematics but recently
due to the availability of empirical data on real-world
networks they have attracted the attention of physics and natural
sciences (see for review 
\cite{ref:ab,ref:dm,ref:n1}). 
Methods of statistical physics, both empirical
and theoretical, have thus begun to play an important role in
this research area. 

The empirical observations of real-networks has had 
a feedback on theoretical development which now 
concentrated on the understanding of the observed
features. For example fat tails in node degree distribution, 
small world effect, degree-degree correlations, or high 
clustering. Two complementary approaches have been developed: 
diachronic, known as growing networks 
\cite{ref:ab,ref:dm,ref:n1},
and synchronic being a sort of statistical mechanics 
of networks \cite{ref:bck,ref:bl1,ref:dms,
ref:bjk1,ref:fdpv,ref:pn1}.

We will discuss here the latter. 
This approach is a natural
extension of Erd\"os and R\'enyi ideas \cite{ref:er,ref:bb}. It is
well suited both for growing (causal) networks for which
nodes' labels reflect the causal order of nodes' attachment 
to the network \cite{ref:kr,ref:bbjk} and for homogeneous 
networks for which nodes' labels can be permuted freely 
in an arbitrary way. 
Here we shall discuss mainly homogeneous networks.
We shall shortly comment on causal networks towards the
end of the paper.

The main aim of the paper is to present
a consistent picture of statistical mechanics of networks.
Some ideas have already been introduced earlier.
They are scattered in many papers and discussed
in many different contexts. We put them together,
add some new material and introduce a guideline
to obtain a self-contained introduction 
to statistical mechanics of complex networks.

The basic concept in the statistical formulation 
is statistical ensemble. Statistical ensemble of networks
is defined by ascribing a statistical weight to every
graph in the given set \cite{ref:bck,ref:bjk1}. Physical quantities are 
measured as weighted averages over all graphs in the ensemble.
The probability of the occurrence of a graph in  
random sampling is proportional to its statistical weight.  
If the statistical weight changes then also the probability of 
occurrence of randomly sampled graphs will change and
in effect different random graphs will be observed. 
The concept of statistical weight is crucial, since
it defines randomness in the system. 
Statistical weight is built out of two ingredients: 
configuration space weight and functional weight. 
The configuration space weight is proportional to the
uniform probability measure on the configuration space 
which tells us how to uniformly choose graphs 
in the configuration space. 
To illustrate the meaning of the 
uniform measure consider an ensemble of Erd\"os-R\'enyi graphs 
with $N$ nodes and $L$ links \cite{ref:er,ref:bb}. 
The configuration space consists of $\binom{\binom{N}{2}}{L}$
graphs with labeled nodes. All those graphs are equiprobable, 
and therefore the configuration space weight
is the same for each graph.
%thus the configuration space weight of each graph in the ensemble
%is the same for each graph. 
It is convenient to choose this weight to
be $1/N!$ since then it can be interpreted as a factor
which takes care of $N!$ possible permutations of nodes' labels.
This factor has the same origin as the corresponding factor in
quantum mechanics for indistinguishable particles and 
it is constant for all graphs in a finite $N$-ensemble. 

We can calculate the entropy of random graphs as
\begin{equation}
S = \ln \frac{1}{N!} \binom{\binom{N}{2}}{L}  .
\label{S0}
\end{equation}
In the limit of large sparse graphs:
$N \rightarrow \infty$ and $\frac{2L}{N} = \alpha = \mbox{const}>2$,
the entropy is an over-extensive function of the system size:
\begin{equation}
S = \frac{\alpha - 2}{2} N \ln N + \dots  ,
\label{S1}
\end{equation}
unlike in standard thermodynamics.

Let us move to weighted graphs.
The idea is to modify the Erd\"os-R\'enyi ensemble by introducing
a functional weight which explicitly depends on graph's topology.
For example, if we choose the functional weight to be a function 
of the number of loops on the graph, we can suppress of favor 
loops of typical graphs in the ensemble.
In a similar way we can
choose statistical weights to control the node degree distribution
to produce homogeneous scale-free graphs \cite{ref:bk}
or to introduce correlations 
between degrees of neighboring nodes \cite{ref:b,ref:n3,ref:bp,ref:d2}.

Classical thermodynamics describes systems in equilibrium for which
the functional weight is given by the Gibbs measure:
$\sim \exp (-\beta E)$, where $E$ is the energy of the system. 
When discussing complex networks it is convenient
to abandon the concept of energy and Gibbs measure
and consider a more general form of statistical weights because 
many networks are not in equilibrium. Indeed, many networks
emerging as a result of a dynamical process like growth
are far from equilibrium \cite{ref:ab,ref:dm,ref:n1}. 
It does not mean though that one cannot
introduce a statistical ensemble of growing networks.
On the contrary, one can for example consider
an ensemble of networks which result of many 
independent repetitions of the growth 
process terminated when the network reaches a certain size.
Such a collection of networks does not describe a thermodynamic
equilibrium. The functional weight can be deduced from the parameters
of the growth process but of course it has nothing to do with 
the Gibbs measure.

In fact, many real-world networks result from a combination 
of a growth process and some thermalization processes.
For example, the Internet grows but at the same time it 
continuously rearranges. The latter process introduces
a sort of thermalization. Today the growth has probably still
larger influence on the topology of the underlying network 
but in the future the growth may slow down due to saturation 
and then equilibration processes 
resulting from continuous rewirings will take over. 
Similarly all evolutionary networks emerge from a
growth mixed with a sort of thermalization related to the continuous network rearrangement.
Therefore it is convenient to have a formalism which can
extrapolate between the two regimes in a flexible way. 
The approach which we propose here is capable of 
modeling functional properties of networks by choosing 
an appropriate functional weight. 

Let us return to the configuration space weight. 
As we mentioned this weight is equivalent
to the uniform probability measure on the configuration
space for which all graphs are equiprobable. It is
a very crucial part of the construction of the ensemble 
to carefully specify what one means by equiprobable graphs. 
Consider first graphs with $N$ nodes.
There are at least two natural candidates for the uniform 
measure in such a set of graphs.
Since one is interested in shape (topology) of graphs one
can define all shapes to be equiprobable. Alternatively one
can introduce labels for nodes of each graph to obtain a set 
of labeled graphs and then one can define all labeled graphs 
to be equiprobable. The two definitions give two different 
probability measures since the number of ways 
in which one can label graph nodes depends
on graph's topology and thus the probability of occurrence of 
a given graph will depend on its topology too.
It turns out that the latter definition is more natural.
As we have seen above this definition leads to 
Erd\"os-R\'enyi graphs. So we stick to this definition 
and from here on we shall ascribe to each
labeled graph the configuration space weight $1/N!$ which
is constant in the set of graphs of size $N$.

The situation is more complex if one considers pseudographs
that is graphs which have multiple connections (more than one
link between two nodes) or self-connections (a link having the same node
at its endpoints). In this case one can also label links 
and ascribe the same statistical weight to each fully labeled graph.
For this choice the statistical weight of each graph is equal
to the symmetry factor of Feynman diagrams generated in the Gaussian
perturbation field theory \cite{ref:bck}.

The paper is organized as follows. In the next section
we will recall some basic definitions. Then we will discuss
Erd\"os-R\'enyi graphs in the context of constructing
statistical ensemble and later we will generalize the
construction to weighted homogeneous graphs.
After this we will describe Monte-Carlo algorithms to 
generate graphs for canonical, grand-canonical and 
micro-canonical ensembles and discuss their representation in terms
of adjacency matrices. A section will be devoted to pseudographs.
In the last section we will shorty summarize the paper.

\section{Definitions}
\begin{figure}[ht]
\psfrag{1}{$1$} \psfrag{2}{$2$} \psfrag{3}{$3$} \psfrag{4}{$4$} \psfrag{5}{$5$}\psfrag{6}{$6$}
\includegraphics[width=2.5cm]{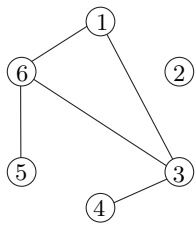}
\caption{An example of simple graph with $N=6,L=5$. 
Vertices without links (like no. 2) are allowed. 
Each vertex can have at most $N\!-\!1$ links.
Positions of vertices in the picture are meaningless.
The only information which matters is connectivity.}
\label{fig:ex0}
\end{figure}

Let us first introduce some terminology. 
Graph is a set of $N$ nodes (vertices) connected by $L$ edges (links).
A graph need not be connected. It may have many disconnected
components including empty nodes (without any link).
If a graph has no multiple or self-connected links we shall
call it simple graph or graph. An example is illustrated in
Fig. \ref{fig:ex0}. Later we shall also discuss graphs with 
multiple- and self-connections. To distinguish them
from simple graphs we shall call them degenerate graphs
or pseudographs. 
One can consider directed or undirected graphs.
Directed graphs are built of directed links while 
undirected of undirected ones. In this paper we
shall discuss undirected graphs but the discussion
can easily be generalized to directed ones as well.
Sometimes we will find it convenient to represent
an undirected link as two oriented links going 
in opposite directions.

A simple graph can be represented by its adjacency matrix which
is an $N\times N$ matrix whose entries $A_{ij}$ are equal 
one if there is a link between vertices $i,j$ or zero otherwise.
Since self-connections are forbidden we have $A_{ii}=0$ on the
diagonal. The adjacency matrix of an undirected graph is also symmetric   
because if there is a link $i\rightarrow j$ ($A_{ij}=1$),
there must be also the opposite one $j\rightarrow i$ 
($A_{ji}=1$).

In this paper we want to construct statistical ensembles 
of homogeneous graphs having desired properties.
We discuss three types of ensembles: ensemble of graphs
with a fixed number of nodes $N$ and varying number of links, 
ensemble with a fixed number of nodes $N$ and a fixed number 
of links $L$, and finally ensemble of graphs 
with a given node degree sequence 
$\{q_1,q_2,\dots, q_N\}$, which we shall call
grand-canonical, canonical and micro-canonical ensembles, respectively.
There are of course many other possibilities like 
for instance ensembles with varying number of nodes, or 
with a fixed number of loops etc, but the three
mentioned above are encountered most frequently. 
To construct a statistical ensemble, for the chosen set of graphs, 
we have to specify statistical weight for each graph in the 
considered set.

In the next section using the Erd\"os-R\'{e}nyi graphs
and binomial graphs we will 
deduce a general logical structure standing behind
the construction of ensembles of homogeneous graphs
and then we will use this structure to introduce ensembles
with an arbitrary functional weight which explicitly
depends on the node degrees. % or the number of loops of the graph.

\section{Statistical ensemble for Erd\"os-R\'{e}nyi random graphs}

\begin{figure}[h]
\includegraphics[width=10cm]{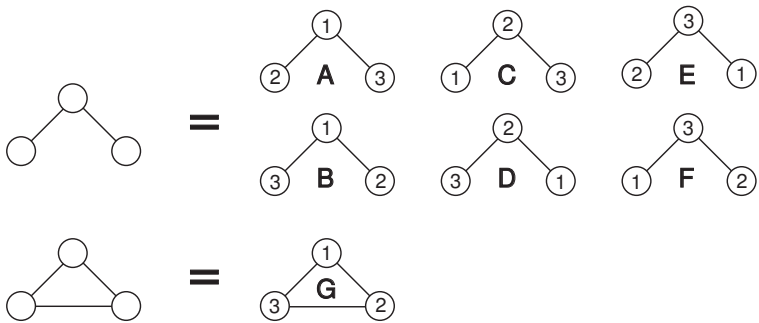}
\caption{Top: the graph on the left can be realized
as three different labeled graphs. A is equivalent to B, C to D 
and E to F. They are equivalent
because they have the same adjacency matrix.
Bottom: triangle-shaped graph has only one realization as labeled graph.}
\label{fig:shape}
\end{figure}

For simplicity,
we start from a well-known model of Erd\"os-R\'{e}nyi's graphs 
\cite{ref:er,ref:bb}. In this classical model one considers
simple graphs with $N$ labeled nodes and $L$ links \footnote{For
simple graphs it is immaterial to label links since each
link is uniquely determined by its end points.}
chosen at random out of all $\binom{N}{2}$ possibilities.
All possibilities are equiprobable and so are the corresponding 
graphs -- understood as graphs whose vertices are labeled.
Usually one is interested in unlabeled graphs
that is in their shape or topology and not
in their labeled version. To explain what is meant by graph's shape 
or topology, let us consider a simple graph 
shown in the upper part of Fig. \ref{fig:shape}. Unlabeled
graph (topology) on the left hand side of the figure 
is represented as labeled graphs on the right hand side.
There are six possible realizations, but
only three of them: A, C, E are distinct. 
B is the same as A since it can be obtained from A 
by a continuous deformation: one can continuously
move the vertex $2$, together with the link attached to it,
to the position of the vertex $3$,
and at the same time the vertex $3$ to the position of the vertex $2$.
Such a continuous deformation does not change graph's connectivity. 
The same holds for pairs: C, D and E, F.
This can also be seen if we take into account the adjacency matrix $\A$.
Both A and B have identical adjacency matrices 
which are different from those for C, D and E, F:
\bq
\A_{\mbox{\scriptsize A}} = \A_{\mbox{\scriptsize B}} = 
\left( \begin{array}{ccc}
	0 & 1 & 1 \\
	1 & 0 & 0 \\
	1 & 0 & 0
\end{array}
\right), \quad
\A_{\mbox{\scriptsize C}} = \A_{\mbox{\scriptsize D}} = 
\left( \begin{array}{ccc}
	0 & 1 & 0 \\
	1 & 0 & 1 \\
	0 & 1 & 0
\end{array}
\right), \quad
\A_{\mbox{\scriptsize E}} = \A_{\mbox{\scriptsize F}} = 
\left( \begin{array}{ccc}
	0 & 0 & 1 \\
	0 & 0 & 1 \\
	1 & 1 & 0
\end{array}
\right) \ .
\eq
Now we remove labels from all graphs in Fig. \ref{fig:shape}
to obtain two unlabeled graphs depicted 
on the left hand side. Although there are three distinct 
adjacency matrices for the upper shape, all of them lead 
to the same connections between vertices (unlabeled graph). 
The graph in the lower line in Fig. \ref{fig:shape}.
can be labeled only in one way \footnote{
At first glance one can think that there are
six ways because one can put labels in six
different ways on a drawing of the triangle. After a while
one can see that they all are identical since they can be
transformed one into another by a transformation which 
does not change connectivity. For example, if it is a drawing
of equilateral triangle one can change labels 123 into 
231 by rotating it by 120$^\circ$.}
which is represented by the following adjacency matrix:
\bq
\A_{\mbox{\scriptsize G}} = 
\left( \begin{array}{ccc}
	0 & 1 & 1 \\
	1 & 0 & 1 \\
	1 & 1 & 0
\end{array}
\right) \ .
\eq
Thus the triangular shape has only one realization as labeled graph.
Furthermore, the upper and lower
graphs are obviously distinct because none of the corresponding
labeled graphs (adjacency matrices) representing the upper graph
is equal to that for the lower graph. 
In this trivial case the difference is in the number of links.
More generally, any two
graphs are distinct if the underlying labeled graphs (adjacency matrices)
cannot be converted one into another by a permutation of
nodes' labels. It is clear that for the graphs in Fig. \ref{fig:shape} 
there is no such a permutation but in general case the comparison 
of graphs may be a complex problem.

\begin{figure}
\includegraphics[width=7cm]{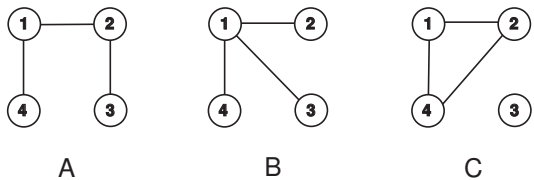}
\caption{Three possible graphs for $N=4,L=3$. 
The number of ways of labeling these graphs is: $n_A=12$, $n_B=4$,
$n_C=4$.}
\label{fig:shape2}
\end{figure}

Let us now apply the ideas sketched above to define an ensemble 
of graphs. As an example we shall consider Erd\"os-R\'{e}nyi graphs 
with $N=4$, $L=3$. It consists of three distinct graphs A, B, C shown
in Fig. \ref{fig:shape2}. 
Now we want to determine the statistical
weight for those graphs.
Adjacency matrices of the underlying labeled graphs 
are essentially different for A, B, C since they cannot be converted
one into another by a permutation of node's labels.
Each graph in Fig. \ref{fig:shape2} has a few possible realizations 
as labeled graph. One can label four vertices in $4!=24$ ways
corresponding to permutations of $1-2-3-4$.
For the graph A, twelve of them give distinct labeled graphs.
For example, the permutation $1-2-3-4$ gives an identical
labeled graph (adjacency matrix) as $4-3-2-1$.
The same kind of symmetry applies for remaining pairs of
permutations. Therefore there are $n_A=12$ labeled graphs for $A$.
Similarly one can find that there are $n_B=4$ labeled graphs for $B$
and $n_C=4$ for $C$. Altogether, there are $n_A+n_B+n_C=20$ labeled
graphs in accordance with $n=\binom{\binom{N}{2}}{L}=20$. In the 
Erd\"os-R\'{e}nyi ensemble labeled graphs are equiprobable,
so the shapes A, B, C have the following probabilities: 
\bq
p_A=\frac{n_A}{n} = \frac{3}{5}, \quad
p_B=\frac{n_B}{n} = \frac{1}{5}, \quad
p_C=\frac{n_C}{n} = \frac{1}{5}. 
\label{pABC}
\eq
These probabilities give frequencies of the occurrence of the 
shapes A, B, C in random sampling. We see that graphs (unlabeled
graphs) are not equiprobable in Erd\"os-Renyi's ensembles. 

Let us denote the statistical weights for A, B, C by $w_A,w_B,w_C$
which are proportional to probabilities 
of configurations in the ensemble. In our case
$w_A:w_B:w_C = p_A:p_B:p_C$. There is a
common proportionality constant in the weights.
It is convenient to choose this constant in such a way that
the weight of each labeled graph be $1/N!$ \footnote{One should
however remember that this constant is an irrelevant proportionality
factor as long as the number of nodes is fixed.}. 
For this choice we have
\bq
w_A=1/2, \quad w_B=1/6, \quad w_C=1/6 ,
\eq
for the graphs in Fig. \ref{fig:shape2}. 
This choice compensates for the increasing factor of permutations $N!$,
when one considers ensembles with varying $N$, and intuitively
removes overcounting coming from summing over permutations 
of indistinguishable node's labels.
However, one should remember that in general
the number of distinct labeled graphs of a graph
is less than $N!$ and therefore the weight
of graph is smaller than $1$. The larger is the symmetry of
a graph topology the smaller is the number of underlying
labeled graphs and thus the smaller is the statistical weight
(see for instance Fig. \ref{fig:shape}). 

The partition function $Z(N,L)$ for the Erd\"os-R\'enyi 
ensemble can be written in the form:
\bq
Z(N,L) = \sum_{\alpha'\in lg(N,L)} \frac{1}{N!} \;\;= 
\sum_{\alpha\in g(N,L)} w(\alpha)  ,
\label{eq:er}
\eq 
where $lg(N,L)$ is the set of all labeled graphs with 
given $N,L$ and $g(N,L)$ is the corresponding set 
of (unlabeled) graphs. The weight $w(\alpha) = n(\alpha)/N!$,
where $n(\alpha)$ is the number of labeled graphs of graph $\alpha$.
We are interested in quantities averaged over the ensemble.
More precisely, we are interested in quantities which depend
on topology of graph and not on node's labels. This means
that if $O(\alpha)$ is such an observable then for any two
labeled graphs $\alpha'_1$ and $\alpha'_2$
of graph $\alpha$ we have $O(\alpha'_1)=O(\alpha'_2)\equiv O(\alpha)$.
The average is defined by
\bq
\left\langle O\right\rangle \equiv 
\frac{1}{Z(N,L)} \sum_{\alpha'\in lg(N,L)} O(\alpha') \frac{1}{N!} \;\;= 
\frac{1}{Z(N,L)} \sum_{\alpha\in g(N,L)} w(\alpha) O(\alpha)  .
\eq
We shall refer to an ensemble with fixed $N,L$ as to
a \emph{canonical ensemble}.
The word ''canonical'' is used here to emphasize that the number of
links $L$ is conserved like the total number of particles in a container
with ideal gas remaining in thermal balance with a source of heat.
The partition function $Z(N,L)$ can be calculated 
by pure combinatorics as we have seen in the introduction.
Now for completeness we derive it using the
adjacency matrix representation of graphs.
The adjacency matrices $\A$ are symmetric,
they have zeros on the diagonal and $L$ unities 
above the diagonal. Thus we have
\bq
Z(N,L) = \sum_{A_{12}} \sum_{A_{13}} \dots \sum_{A_{1N}} \sum_{A_{23}} \sum_{A_{24}} 
\dots \sum_{A_{N-1,N}} \delta\left[L-\sum_{p<r} A_{pr}\right] \; 1/N!  ,
\eq	
where $\delta\left[x\right]=1$ if $x=0$ and zero elsewhere.
%The sums over $A_{ij}$ run over the set $0,1$.
The sums are done over all $A_{ij}=0,1$ for all pairs $i,j: \, 1\leq i<j\leq N$.
Using an integral representation of $\delta\left[ x \right]$ 
and exchanging the order of summation and integration we obtain
the expected result:
\ba
Z(N,L) & = & (1/N!) \frac{1}{2\pi} \int_{-\pi}^{-\pi} dk \; 
e^{ikL} \left( 1+e^{-i k} \right)^{\binom{N}{2}}
= (1/N!) \frac{1}{2\pi} \int_{-\pi}^{-\pi} dk \; e^{ikL} 
\sum_{m=0}^{\binom{N}{2}} \binom{\binom{N}{2}}{m} e^{-ikm} \nonumber \\
& =  & (1/N!) \binom{\binom{N}{2}}{L}  .
\ea
This method can be applied to calculate averages of various quantities.
As an example consider the
node degree distribution $\pi(q)$ which tells us what is 
the probability that a randomly chosen vertex on random graph 
has degree $q$:
\bq
\pi(q) = \left\langle \frac{1}{N} \sum_i \delta \left[q-q_i\right] 
\right\rangle  .
\label{piq}
\eq
By random graph we mean that we average over graphs from 
the given ensemble.
We know that for Erd\"os-R\'enyi graphs $\pi(q)$ 
is a Poissonian distribution in the limit 
of $N\rightarrow\infty$:
\bq
\pi(q) = \frac{\bar{q}^q}{q!} \exp(-\bar{q})  ,
\label{eq:piq}
\eq
where $\bar{q}=2L/N$ is the average vertex degree.
This result can be rederived using the method described above.
Let us look at the degree of a vertex labeled by one. The result
does not depend on the vertex label for homogeneous graphs since
labels have no physical meaning. One can find that
\ba
\pi(q) & = & \frac{1}{Z(N,L)} \frac{1}{N!} 
\sum_{A_{12}} \sum_{A_{13}} \dots 
\sum_{A_{23}} \sum_{A_{24}} \dots \sum_{A_{N-1,N}} 
\delta\left[L-\sum_{p<r} A_{pr}\right] \; 
\delta\left[ q - \sum_{r=2}^N A_{1r} \right] \nonumber \\
& = &  
\binom{\binom{N}{2}}{L-q} \binom{N-1}{q} / \binom{\binom{N}{2}}{L}  .
\label{eq:exact_p}
\ea
which in the limit $\bar{q}=\mbox{const},\,N\rightarrow\infty$ gives
Eq. (\ref{eq:piq}) as expected.

So far we have discussed the canonical ensemble of Erd\"os-R\'{e}nyi
graphs with $N,L$ fixed.  Erd\"{o}s and R\'{e}nyi
introduced also a related model called {\it binomial model}
where the number of nodes $N$ is fixed but the number of links $L$
is not fixed {\em a priori}. One starts from $N$ empty vertices and connects
every pair of vertices with a probability $p$.
In this statistical ensemble the probability of obtaining 
a labeled graph with given $L$ is $P(L)=p^L (1-p)^{\binom{N}{2}-L}$.
Thus the partition function is
\ba
Z(N,\mu) & = & \sum_L \sum_{\alpha\in lg(N,L)} \frac{1}{N!} P(L(\alpha)) 
= (1-p)^{\binom{N}{2}} \sum_L \left(\frac{p}{1-p}\right)^L 
\sum_{\alpha\in lg(N,L)} \frac{1}{N!} \nonumber \\
& \propto & \sum_L \exp (-\mu L) \; Z(N,L)
\propto \sum_L \exp (-\mu L + S(N,L))  ,
\label{grand}
\ea
where $\frac{p}{1-p}\equiv \exp(-\mu)$ or $\mu = \ln \frac{1-p}{p}$,
and the entropy $S(N,L)$ is given by Eq. (\ref{S0}).
We skipped an $L$-independent factor
in front of the sum in the second line substituting equality
by proportionality sign.
The weight of labeled graphs is $w(\alpha)=1/N! \; \exp(-\mu L(\alpha))$,
where $\mu$ is a constant which can be interpreted as chemical potential 
for links in the \emph{grand-canonical} ensemble (\ref{grand}).
One can calculate the average number of links or
its variance as derivatives of the grand-canonical partition
function with respect to $\mu$: 
$\langle L \rangle = -\partial_\mu \ln Z(N,\mu)$
and $\langle L^2 \rangle -\langle L \rangle^2
= \partial^2_\mu \ln Z(N,\mu)$.
The sum of states can be done exactly:
\bq
	Z(N,\mu) = \sum_{L=0}^{\binom{N}{2}} e^{-\mu L} \frac{1}{N!} \binom{\binom{N}{2}}{L} = 
  \frac{1}{N!} (1+e^{-\mu})^{\binom{N}{2}}  .
\eq
It is easy to see that for fixed finite $\mu$ the average
number of links behaves as $N^2$ or more precisely as
\bq
\langle L \rangle = p \frac{N(N-1)}{2} = 
\frac{1}{1 + e^\mu} \frac{N(N-1)}{2}   .
\label{Lp}
\eq
Thus for $N\rightarrow\infty$ the graphs become dense. 
The mean value of node degree 
$\langle \bar{q} \rangle =2\langle L \rangle /N$ increases to infinity.
The situation changes when $\mu$ goes to infinity with increasing $N$.
This happens in particular if the probability $p$
scales as $p \sim 1/N$ since then $\mu$ behaves as 
$\mu \sim \ln N$. In this case 
$L$ is proportional to $N$ (\ref{Lp}) and
both the terms $\mu L$ and $S$ 
in the exponent of Eq. (\ref{grand}) behave as $N\ln N$ 
and compensate each other. The corresponding graphs 
become sparse and the mean node degree
$\langle \bar{q} \rangle =2\langle L \rangle /N$ is now finite.
The situation in which $\mu$ scales as $\ln N$
is very different from the situation known from classical 
statistical physics, where such quantities like chemical potential 
$\mu$ are intensive and do not depend on system size $N$ 
in the thermodynamic limit $N\rightarrow\infty$. 

The difference between canonical and grand-canonical ensembles
gradually disappears in the large $N$ limit \cite{ref:bck,ref:dms}. 
For canonical ensemble or sparse graphs the node degree 
$\bar{q} = 2L/N = \alpha$ is kept constant 
when $N\rightarrow \infty$ while in grand-canonical one it may fluctuate
around the average $\langle \bar{q} \rangle  = 
2\langle L\rangle /N = \alpha$. However, the magnitude of
fluctuations around the average disappears in the large $N$ limit
since
\bq
\langle L^2 \rangle - \langle L\rangle^2 = 
\binom{N}{2} \frac{e^{-\mu}}{(1+e^{-\mu})^2}  ,
\eq
and $\Delta q = 
\sqrt{\langle L^2 \rangle - \langle L\rangle^2} / \langle L \rangle
\sim 1/N \rightarrow 0$,
so effectively the system selects graphs with $\bar{q} = \alpha$.

Sometimes one also considers an ensemble of graphs
with a predefined node degree sequence $\{q_1,q_2,\dots,q_N\}$.
We shall call it \emph{micro-canonical}.
Again, in the simplest case one assumes that all
labeled graphs are equiprobable in this ensemble. Properties
of random graphs in such an ensemble 
strongly depend on the degree sequence.

\section{Weighted homogeneous graphs}
In the previous section we described 
ensembles for which all labeled graphs 
had the same statistical weight. Random graphs 
in such ensembles have well known 
properties. It turns out, however, that most
of these properties do not correspond to those
observed for real world networks. One needs a more
general set-up to define an ensemble of complex random 
networks. Such a set-up can be introduced as follows.
One considers the same set of graphs as in Erd\"os-R\'enyi model
but one ascribes to each graph a different statistical weight. 
In other words, one chooses a probability measure 
on the set of graphs which differs from the uniform measure.
In the generalized ensemble, each graph 
in addition to the configuration space weight $1/N!$ 
has a functional weight $W(\alpha)$.
% which
%is responsible for the frequency of random graphs having
%a certain property. 
For homogeneous random graphs this weight 
is assumed to depend only on graph topology. This means
that the weight does not depend on nodes' labels:
if $\alpha'_1$ and $\alpha'_2$ are labeled graphs
of $\alpha$ then $W(\alpha'_1)=W(\alpha'_2)\equiv W(\alpha)$.
The partition function for a weighted canonical ensemble
reads
\bq
Z(N,L) = \sum_{\alpha'\in lg(N,L)} (1/N!)\, W(\alpha') \;\;= 
\sum_{\alpha\in g(N,L)} w(\alpha) W(\alpha)  ,
\label{eq:can_g}
\eq
where $w(\alpha)$ is the same factor $w(\alpha) = n(\alpha)/N!$
as before (\ref{eq:er}), being just the ratio of the number of 
labeled graphs $\alpha'$ of $\alpha$ (obtained by 
permutations of nodes' labels giving distinct adjacency matrices)
and the number of all nodes' labels permutations $N!$.
For Erd\"os-R\'enyi graphs the functional weight is $W(\alpha)=1$.

The simplest non-trivial example is 
a family of product weights $W$:
\bq
W(\alpha) = \prod_{i=1}^N p(q_i)  ,
\label{eq:prod}
\eq
where $p(q)$ is a positive function depending on 
one node degree $q$.
This functional weight does not introduce
correlations between node degrees. We shall refer to
random graphs generated by this partition function as
\emph{uncorrelated networks}. One should however remember
that the total weight does not entirely factorize because the
configuration space weight $w(\alpha) = n(\alpha)/N!$ written
as a function of node degrees $w(q_1,q_2,\dots,q_N)$ does
not factorize. There is also another factor which prevents
the model from the full factorization and independence
of node degrees, namely this is the total number of links 
$2L = q_1+q_2+\dots + q_N$ which for given $L$ and $N$ introduces
correlations between $q_i$'s. For example, if one of $q_i$'s is
large, say of order $2L$, then the remaining ones have to be small
in order not to violate the constraint on the sum. 
For a wide class of weights $p(q)$ one can however show
that in the large $N$ limit the probability that a randomly 
chosen graph has degrees $q_1,\dots,q_N$ approximately factorizes:
\bq
\pi(q_1,\dots,q_N) \sim \prod_{i=1}^N \pi(q_i)   .
\label{fact}
\eq
For large $N$, the node degree distribution $\pi(q)$ (\ref{piq}),
that is the probability that a random node on random graph has 
degree $q$ \cite{ref:bck,ref:bjk1,ref:bl1}, can be approximated by
\bq
\pi(q) = \frac{p(q)}{q!} \exp(-Aq-B)  ,
\label{piqAB}
\eq
where parameters $A,B$ are determined from the conditions 
for the normalization $\sum_q \pi(q)=1$ and for the average 
$\sum_q q \pi(q) = \bar{q} \equiv 2L/N$.
For example, for $p(q)=1$ which corresponds to
Erd\"os-R\'{e}nyi graphs one finds $A = -\ln \bar{q} = \ln 2L/N$
and $B = \bar{q} = 2L/N$, therefore $\pi(q)$ is given by the Poissonian
from Eq. (\ref{eq:piq}).  

Since the node degree distribution $\pi(q)$ 
for weighted graphs (\ref{eq:prod}) depends on $p(q)$, one can
choose the latter to obtain a desired form of 
the node degree distribution $\pi(q)$. Let $\pi(q)$ 
be a desired node degree distribution such that
\begin{equation}
\sum_q \pi(q)=1 \ , \quad \bar{q} = \sum q \pi(q) .
\end{equation}
If we choose the weight (\ref{eq:prod}) with
\begin{equation}
p(q) = q! \pi(q)
\end{equation}
in canonical ensemble with $N$ nodes
and $L$ links, in the limit of $N\rightarrow \infty$ and $2L/N = \bar{q}$
we obtain homogeneous random graphs 
with this node degree distribution.
In this case the constants $A$, $B$ from Eq. (\ref{piqAB}) vanish
automatically: $A=B=0$.
In particular by an appropriate choice of $p(q)$ we
can generate scale free graphs with the node degree Barab\'{a}si - Albert distribution \cite{ref:bamodel}:
$\pi(q) = \frac{4}{q(q+1)(q+2)}$ for $q=1,2,\dots$ and $\pi(0)=0$
as an ensemble of graphs $L=N$, $\bar{q} = 2$
with $p(q) = q! \frac{4}{q(q+1)(q+2)}$ for $q=1,2,\dots$ and $p(0)=0$.
However, for finite $N$ the node degree distribution 
$\pi(q)$ deviates from the limiting shape due to finite size corrections,
which are particularly strong for fat tailed distributions $\pi(q) \sim q^{-\gamma}$.
The maximal node degree 
scales as $q_{max}\sim N^{1/(\gamma-1)}$ for $\gamma\ge 3$ and 
as $q_{max} \sim N^{1/2}$ for very fat tails: $2 < \gamma < 3$ 
\cite{ref:bk,ref:bpv} as a result of structural constraints
which also lead to the occurrence of correlations between node degrees.  

One can define more complicated weights than those given by 
Eq. (\ref{eq:prod}). A natural candidate for networks
with degree-degree correlations is the following
weight \cite{ref:bl1,ref:b}:
\bq
W(\alpha) = \prod_{l=1}^L p(q_{a_l},q_{b_l})  ,
\eq
where the product runs over all links of the graph,
and the weight $p(q_a,q_b)$ is a symmetric function
of degrees of nodes at the end points of the link.
One can choose this function to favor assertive
or disassertive behavior \cite{ref:bl1,ref:b,ref:n3,ref:bp,ref:d2}. 
In a similar way one can
introduce probability measures on the set of graphs
which mimic some other functional properties of real 
networks, like for example higher clustering 
\cite{ref:n2,ref:bjk2,ref:bjk3,ref:d1,ref:pn2}. 
One can do this in micro-canonical, canonical, grand-canonical
or any other ensemble. This is just the most general
set-up to handle homogeneous networks.

\section{Monte-Carlo generator of homogeneous networks}

Erd\"os-R\'enyi graphs are exceptional in the sense
that one can calculate for them almost all quantities of interest
analytically. This is not the case for weighted networks.
Various methods have been proposed for generating random graphs 
\cite{ref:k}.
In this section we will describe a Monte-Carlo method
which allows one to study a wide class of random weighted
graphs experimentally by a sort of numerical experiments.
The basic idea behind this type of experiments is to sample
the configuration space of graphs with the probability 
proportional to the statistical weight or in other
words to generate graphs with a desired probability.
Again, the Erd\"os-R\'enyi graphs are exceptional 
because one can generate them one by one independently of 
each other. This is just because they are equiprobable. 
For weighted graphs the situation is not that easy since
there are no efficient algorithms to pick up
an element from a large set with the given probability.
The naive algorithm which relies on picking up an element
uniformly and then accepting it with the given
probability has a very low acceptance rate.
Therefore one has to use another idea. We will
describe below how to generate graphs using 
dynamical Monte-Carlo technique.

The idea is to run a random walk process 
in the set of graphs which visits configurations with 
a frequency proportional to their 
statistical weight. Mathematically, this means
that one has to invent a stationary
Markov chain (process) for which the stationary distribution
is proportional to the statistical weights of graphs:
$\sim W(\alpha)/Z$. 

The Markov chain is defined by
transition probabilities $P(\alpha\rightarrow \beta)$
that the random walker will go in one step 
from a configuration (graph) $\alpha$ to $\beta$. 
The probabilities are stored in a transition matrix $\pp$:
$P_{\alpha\beta} \equiv P(\alpha\rightarrow \beta)$
which is also called Markov's matrix. For a stationary process,
the transition matrix $\pp$ 
is constant during the random walk.
Random walk is initiated from a certain graph 
$\alpha_0$ and then elementary steps are
repeated producing a sequence (chain) of graphs 
$\alpha_0 \rightarrow \alpha_1 \rightarrow  \alpha_2 \rightarrow \dots \ $.
The probability $p_\beta(t+1)$ that
a graph $\beta$ will be generated in the $(t+1)$-th step 
of the Markov process can be calculated as
\bq
p_\beta(t+1) = \sum_\alpha p_\alpha(t) P_{\alpha \beta}   .
\eq
The last equation can be written as
\bq
\p (t+1) = \pp^{\tr} \p (t)  ,
\label{interp}
\eq
where $\tr$ denotes transposition, and $\p$ is a
vector of elements $p_\alpha$. One should note that the stationary
state: $\p(t+1) = \p(t)$ corresponds to a left eigenvector 
of $\pp$ to the eigenvalue \footnote{One can show that all eigenvalues 
of a Markov transition matrix lie inside or on the unit 
circle $|\lambda| \le 1$.}
$\lambda=1$.
If the process is ergodic, which means that 
any configuration can be reached by a sequence of transitions
starting from any initial configuration,
and if the transition matrix fulfills the detailed balance
condition:
\bq
W_\alpha P_{\alpha \beta} = 
W_\beta P_{\beta \alpha} \quad \forall \alpha,\beta  ,
\label{eq:db}
\eq
then the stationary state can be shown to approach the
desired distribution:
$p_\alpha(t) \rightarrow W_\alpha/Z$ for $t \rightarrow \infty$.
We used a short-hand notation $W_\alpha$ for $W(\alpha)$.
In other words, when the length of the Markov chain becomes
infinite the probability of occurrence of graphs 
in the Markov chain becomes proportional to their
statistical weights and becomes independent of the initial 
configuration. Therefore the average over graphs
generated in this Markovian random walk is a good
estimator of the average over the weighted ensemble.
The price to pay for generating graphs in this way 
is that the consecutive graphs in the Markov chain may 
be correlated with each other. 
Therefore one has to find a minimal number of steps for which
one can treat measurements on such graphs as independent.

One should note that the only
characteristics of the Markov process which
has a physical meaning from the point of view of the
simulated ensemble is the stationary distribution.
All other dynamical properties of the random walk 
which are encoded in the form of transition matrix
$P(\alpha\rightarrow\beta)$
are irrelevant. Many different 
transition matrices $\pp$
may have the same stationary distribution. 
Indeed, many of them fulfill the detailed balance condition for 
given weights $W_\alpha$ (\ref{eq:db}). The best known
choice of $\pp$ is
\bq
P_{\alpha\beta} = 
\min\left\{1,\frac{W_\beta}{W_\alpha}\right\}  .
\label{metrop}
\eq
This choice is quite general and can be used in many
different contexts. It is called Metropolis
algorithm. 
For the current configuration $\alpha$ one proposes a change 
to a new configuration $\beta$ which differs slightly from
$\alpha$ and then one accepts it with
the Metropolis probability (\ref{metrop}). One
repeats this many times producing a chain of configurations.
The proposed modifications should not be too large
since then the acceptance rate would be small. Therefore
the algorithm makes only small steps (moves) in the configuration
space which form a sort of weighted random walk.

\section{Monte-Carlo generator of canonical ensemble}
\begin{figure}[h]
\psfrag{i}{$i$} \psfrag{j}{$j$} \psfrag{k}{$k$}
\includegraphics[width=13cm]{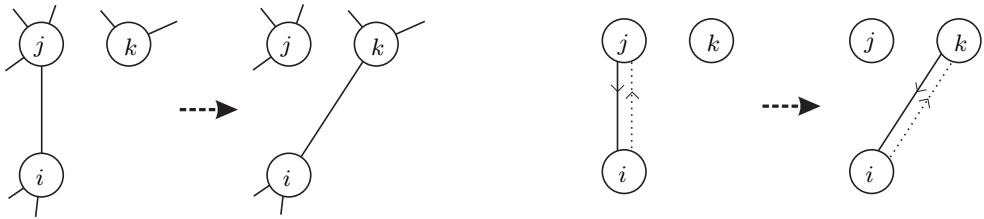}
\caption{The idea of rewiring: a random link (dotted line) 
is rewired from vertex $j$ to a random vertex $k$ (left hand side).
Alternatively (right hand side) a random, oriented link 
(dotted line) is rewired from vertex of its end $j$ 
to a random vertex $k$. The opposite link 
$j\rightarrow i$ is simultaneously rewired.}
\label{fig:tmove}
\end{figure}

Now, we want to apply this method to generate Erd\"os-R\'{e}nyi
graphs. Let us begin with the canonical ensemble with $N,L$ fixed. 
A good candidate for elementary transformation of graph 
is rewiring of a link as shown in Fig. \ref{fig:tmove},
because it does not change $N$ and $L$.
As mentioned before it is convenient to introduce 
a representation in which each undirected link is represented
by two directed links. 
%\begin{figure}[ht]
%\includegraphics[width=12cm]{tmove0.eps}
%\caption{The idea of rewiring: a random link (dotted line) 
%is rewired from vertex $j$ to a random vertex $k$.}
%\label{fig:tmove0}
%\end{figure}
The rewiring is done in two steps \cite{ref:bck}. First we 
choose a directed link $ij$ and a vertex $k$ at random.
Then we rewire the link $ij$ to $ik$. If there is already
a link between $i$ and $k$ or if the vertex $k$ coincides with $i$,
we reject the rewiring since it would 
otherwise lead to a multiple- or self-connection. 
One should note that the result of rewiring $ij$ 
is not the same as of rewiring $ji$. 
The move is accepted with the Metropolis probability.
For the canonical ensemble of Erd\"os-R\'enyi
graph this probability is equal to one
since functional weights are $W_\alpha=W_\beta=1$ in Eq. (\ref{metrop}).

Let us see how rewiring transformations work in practice.
Consider as an example the set of graphs shown in Fig. \ref{fig:shape2}.
If we pick up the link $3-2$ in the graph A 
and rewire it to $3-1$, we will obtain the graph B.
If we rewire the link $2-3$ to $2-4$, we will get the graph C.
So using the procedure of rewiring showed in Fig. \ref{fig:tmove} 
we can obtain every graph in the ensemble.
The rewiring transformation is ergodic in this set of graphs.

To summarize, our procedure of generating graphs in this training 
ensemble looks as follows. We construct an arbitrary graph having
$N=4$ nodes and $L=3$ links to initiate the procedure
and then we repeat iteratively rewirings for
randomly chosen links and vertices. The only restriction is
that the rewirings do not produce self- or multiple-connected
links. We keep on repeating until we obtain ''thermalized graphs''. 
Then we can begin measuring quantities on the generated sequence 
of random graphs. 

\begin{figure}[ht]
\includegraphics[width=3cm]{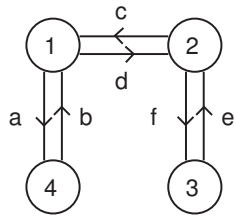}
\caption{The representation of graph A in Fig. \ref{fig:shape2} 
as directed graph. }
\label{fig:tmove_ex}
\end{figure}

Let us check that the described Monte-Carlo procedure indeed
generates graphs with the expected probabilities (\ref{pABC}).
Let us calculate the Markovian matrix $\pp$ for the
rewiring procedure in this ensemble. First we calculate the
transition probability from A to B. The graph
A can be converted into B in one step if
we rewire the link ''b'' in Fig. \ref{fig:tmove_ex} to the vertex 2,
or alternatively the link ''e'' to the vertex 1. We see that for this
change we can choose two out of six links and one of four vertices 
to obtain the desired change. Thus the probability of choosing 
links is $2/6$ and of choosing correct vertex is $1/4$, so
the total probability is $P(A\rightarrow B)=2/6 \cdot 1/4 = 2/24$.
Let us now calculate $P(A\rightarrow C)$.
To obtain $C$ from $A$ we have to rewire
''a'' to 3 or ''f'' to 4. Thus $P(A\rightarrow C)=2/6 \cdot 1/4 = 2/24$.
We can find $P(A\rightarrow A)$ from the condition:
$P(A\rightarrow A)+P(A\rightarrow B)+P(A\rightarrow C)=1$.
This yields $P(A\rightarrow A)=20/24$.
Repeating the calculations for the remaining cases we find:
$P(B\rightarrow A)=6/24$,
$P(B\rightarrow B)=18/24$, $P(B\rightarrow C)=0$ and
$P(C\rightarrow A)=6/24$, $P(C\rightarrow B)=0$, 
$P(C\rightarrow C)=18/24$. The results can be collected
in a transition matrix:
\bq
\pp = \frac{1}{24} \left[\begin{array}{ccc}
	20 & 2 & 2 \\
	6 & 18 & 0 \\
	6 & 0 & 18
\end{array}\right].
\eq 
We can now determine the stationary probability distribution 
of the Markov process as a left eigenvector to the eigenvalue
one of the transition matrix $\pp$. We obtain
$p_A:p_B:p_C = 3:1:1$ in agreement with Eq. (\ref{pABC}). 
This is not surprising since
$\pp$ satisfies the detailed balance condition (\ref{eq:db})
and the corresponding changes are ergodic.

We have checked above by explicit calculation
that the algorithm gives correct weights of
Erd\"os-R\'enyi graphs for $N=3,L=4$.
One can give a general argument that for any $N,L$
the algorithm generates labeled graphs which are
equiprobable. Suppose that we have a certain labeled 
graph $\alpha$ and want to get $\beta$ by rewiring $ij$ to $ik$. 
(see Fig. \ref{fig:tmove}). The total probability
$P(\alpha\rightarrow\beta)$ can be written
as a product of two factors: the probability $P_c$ of choosing 
a particular candidate for a new configuration and 
the probability $P_a$ of accepting it.
Because we choose a link $i\rightarrow j$ from $2L$ possible 
directed links and a vertex $k$ from $N$ vertices we 
have $P_c = 1/(2LN)$. Inserting
$P(\alpha'\rightarrow\beta') = 1/(2LN) \, 
P_a(\alpha'\rightarrow\beta')$ in the Eq. (\ref{eq:db})
and similarly for $\alpha'\leftrightarrow \beta'$ we get
\bq
w_{\alpha'} P_a(\alpha'\rightarrow\beta') = 
w_{\beta'} P_a(\beta'\rightarrow\alpha')  .
\label{eq:can1}
\eq
But $w_{\alpha'}=1/N!$ for all labeled graphs, 
thus $P_a(\alpha'\rightarrow\beta')=P_a(\beta'\rightarrow\alpha')$.
This means that every move should be accepted unless it
violates the multiple- or self-connections constraints.
The rejection does not change the frequency of the occurrence of
simple graphs but only restricts the space of sampled graphs
to what we need. The weights of (unlabeled) graphs $\alpha$
are in this case $w(\alpha)= n(\alpha)/N!$ where $n(\alpha)$ 
is the number of distinct graphs of $\alpha$.

\begin{table}[h]
\begin{center}
\begin{tabular}{|c|c|c|c|c|c|c|}
\hline
Graphs & \multicolumn{6}{|c|}{\includegraphics[width=7cm]{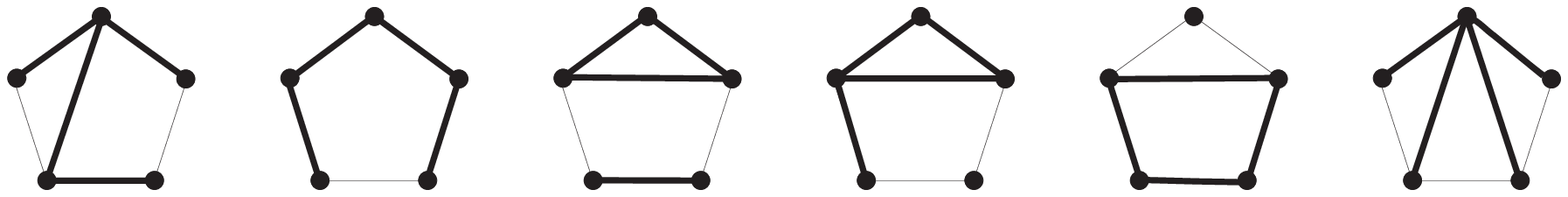}}	\\
\hline
$w_\alpha$ & 1/2 & 1/2 & 1/12 & 1/2 & 1/8 & 1/24 \\
$p_\alpha$ & 0.286 & 0.286 & 0.048 & 0.286 & 0.071 & 0.024 \\
\hline
\multicolumn{7}{|c|}{Simulations} \\
\hline
R & \multicolumn{6}{|c|}{} \\
\hline
10 & 0.285(1) & 0.286(1)   &   0.048(1)   &   0.286(1)   &   0.071(1)   &   0.024(1)   \\
5 & 0.286(1) & 0.285(1)   &   0.047(1)   &   0.286(1)   &   0.071(1)   &   0.024(1)   \\
2 & 0.285(1) & 0.285(1) &   0.047(1)   &   0.286(1)   &   0.072(1)   &   0.024(1)   \\
1 & 0.284(1) & 0.286(1) &   0.048(1)   &   0.284(1)   &   0.072(1)   &   0.025(1)   \\
\hline
\end{tabular}
\caption{Theoretically calculated weights $w_\alpha$
of graphs in the canonical ensemble $N=5,L=4$ are
normalized to ensure probabilistic interpretation:
$p_\alpha = w_\alpha/\sum_\beta w_\beta$,
and compared with the experimental frequencies in the Markov chain
using algorithm of rewiring. The results differ by the number
$R$ of rewirings between consecutive measurements. }
\label{tab1}
\end{center}
\end{table}
Let us numerically test the algorithm.
In table \ref{tab1} we compare the weights calculated analytically
and computed from Monte-Carlo generated graphs
for $N=5,L=4$. There are six different graphs in this ensemble.
We generated $10^6$ graphs. Each of 
them was obtained from the previous one by $R$
rewirings, or more precisely by $R$ attempts of rewiring \footnote{
Even if a rewiring is rejected we count it in since it
corresponds to the transition $P(\alpha \rightarrow \alpha)$
from graph to the same graph.}.

As we can see in table \ref{tab1},
the frequency of occurrence of each graph is in
an excellent agreement with the expected weights. 
The results do not depend on the separation $R$ 
between the measurements.
In the chain of $10^6$ graphs, each graph in this ensemble
is produced many times. For larger ensembles 
the algorithm would not be able to visit all graphs since
the number of graphs is very large (\ref{S1}). In this
case the algorithm would
choose only those graphs which are most representative. To make sure
that the algorithm has reached the stationary distribution
one should start a couple of random walks from different corners of the
configuration space and run the algorithm so long as the
statistical properties of graphs generated in all the random 
walks become identical.

Generalization of the algorithm to a weighted ensemble 
is straightforward. We insert
the statistical weights $W_\alpha$ of this ensemble
into the Metropolis formula (\ref{metrop}). 
Consider in particular a product weight (\ref{eq:prod}). 
We choose a link $ij$ and a vertex $k$ 
on the current configuration $\alpha$ at random 
and attempt to rewire the link to $ik$ to obtain 
a new configuration $\beta$.
The change is accepted with the probability
\bq
P_a(\alpha\rightarrow\beta) = 
\min\left\{1,\frac{W_\beta}{W_\alpha}\right\} = 
\min\left\{1,\frac{p(q_j-1)p(q_k+1)}{p(q_j)p(q_k)}\right\}  .
\eq
The degrees $q_j,q_k$ are taken from $\alpha$.
Clearly the rewiring changes the degrees
$q_j\rightarrow q_j-1$, $q_k\rightarrow q_k+1$,
and leaves the degrees of remaining nodes intact. 
The ratio $W_\beta/W_\alpha$ can be calculated for
any form of statistical weights, so the algorithm is
general.

\section{Monte-Carlo generator of grand-canonical ensemble}
The rewiring procedure described in the previous section
does not change $N$ and $L$. If we want to simulate graphs
from a grand-canonical ensemble for which $L$ is variable,
we have to supplement the set of elementary transformations
in the algorithm by transformations which change the number of links.
We can introduce two mutually
reciprocal transformations: adding and deleting a link.
Both they preserve the number of nodes $N$ but change the
number of links: $L \rightarrow L\pm 1$. 
The two transformations must be carefully balanced.
On a given graph $\alpha$ we have to choose one of them.
Let the link addition be selected with the probability
$p_+$ and the removal with $p_-$. Once the move is
selected we have to choose a link-candidate for which the
move is to be applied. It is convenient to split the total 
transition probability into three factors: 
\bq
P(\alpha\rightarrow \beta) = p_\pm P_c(\alpha\rightarrow \beta) 
P_a(\alpha\rightarrow\beta)   ,
\eq
where $p_\pm$ stands for one of $\{p_-,p_+\}$,
$P_c(\alpha\rightarrow \beta)$ for the probability of choosing
a candidate configuration for the change
and $P_a(\alpha\rightarrow \beta)$ for the probability 
of accepting the move. Let $\alpha$ and $\beta$ be two graphs
which differ by a link which is present on $\beta$ 
but absent on $\alpha$: $L(\alpha)=L(\beta)-1$.
The transition probability for adding
a link to $\alpha$ has to be balanced with the probability of
removing the link from $\beta$. 
In order to add a link to $\alpha$ we have to choose two vertices
to which the addition of a link is attempted.
The probability of choosing a given pair of vertices,
if we choose two vertices independently, is
$P_c(\alpha\rightarrow\beta)=2/N^2$. Thus the total
probability of this move is
\bq
P_{\alpha\beta}=
P(\alpha\rightarrow\beta) = p_+ \, \frac{2}{N^2} \,
P_a(\alpha\rightarrow\beta) .
\eq
In the reciprocal transformation we have to choose this
link among all links. The probability of choosing one among $L$ links
is $P_c(\beta\rightarrow\alpha)=1/L_\beta=1/(L_\alpha+1)$.
Thus the total probability of this move is
\bq
P_{\beta\alpha}= P(\beta\rightarrow\alpha) = p_- \, \frac{1}{L_\beta} \,
P_a(\beta\rightarrow\alpha) .
\eq
Now we have to insert the last two equations to the 
detailed balance condition which for the grand-canonical
ensemble additionally includes the factor $e^{-\mu L}$:
\bq
W_\alpha e^{-\mu L_\alpha} P_{\alpha \beta} = 
W_\beta e^{-\mu L_\beta} P_{\beta \alpha} \ .
\label{eq:db_gc}
\eq
Using this we can calculate the ratio
\bq
\frac{P_a(\alpha\rightarrow\beta)}{P_a(\beta\rightarrow\alpha)} =
\exp(-\mu) \, \frac{p_-}{p_+} \, 
\frac{N^2}{2L_\beta} \, \frac{W_\beta}{W_\alpha} .
\eq
If one chooses the same number of attempts for adding
and removing a link: $p_+=p_-$, then the ratio
$p_+/p_-=1$ will disappear from the last equation
and the acceptance probabilities for 
adding or removing a link in the Metropolis algorithm
will read
\bq
P_a(\alpha\rightarrow\beta) = 
\min\left\{1, \exp(-\mu) \, 
\frac{N^2}{2(L_\alpha+1)} \, \frac{W_\beta}{W_\alpha} \right\} ,
\label{eq:gcan_rown}
\eq
and
\bq
P_a(\beta\rightarrow\alpha) = 
\min\left\{1, \exp(+\mu)  \,
\frac{2L_\beta}{N^2} \, \frac{W_\alpha}{W_\beta}\right\} ,
\label{eq:gcan_rown2}
\eq
respectively. 
As before if we want to produce only simple graphs we must
have an additional condition which eliminates moves leading 
to self- or multiple connections.
The algorithm is complete. One should note
that there is no reason to do additional rewirings because 
a rewiring of a link 
$ij$ to a link $ik$ is equivalent to removing
the link $ij$ and adding $ik$.

In principle one could propose other algorithms. For example,
one could consider a modified algorithm in which
the move removing a link is done in a different way. 
Instead of picking up a link as a candidate,
one could pick up two vertices at random, and then remove
a link if there is any between them.
The probability of choosing a pair of vertices would
be $2/N^2$ and it would cancel with the identical factor
for the probability of choosing candidates in the move 
adding a link. The fractions
$N^2/2L$ and $2L/N^2$ would in this case disappear from equations
 	(\ref{eq:gcan_rown}) and (\ref{eq:gcan_rown2}). 
The two algorithms of course generate 
the same ensemble. However, the modified algorithm would have
much worse acceptance rate for sparse networks since 
the chance that there is a link between 
two randomly chosen vertices on a sparse graph 
is very small. Most of the chosen pairs of vertices 
are not connected by a link 
and therefore the algorithm will do nothing since 
there is no link to remove.

This problem is absent for the
algorithm which we described previously since in that case
only existing links are chosen as candidates for removal.
One can easily estimate that the probability of accepting
a link removal (\ref{eq:gcan_rown2}) is not very small. 
Indeed, even for sparse graph the factor $e^{\mu} 2L/N^2$ 
in Eq. (\ref{eq:gcan_rown2}) is of order unity. In this 
case both $\exp(\mu)$ and $L$ for large $N$ grow proportionally
to $N$ and their product balances the factor
$N^2$ in the denominator. The
algorithm has a finite acceptance rate which does not
vanish when the system size grows.

As an exercise, let us consider an example of unweighted 
($W_\alpha=1$) graphs with $N=3$. This ensemble consists 
of four graphs shown in table \ref{tab2}. 
Their statistical weights can be easily found to be 
$1/3!, \, 3e^{-\mu}/3!, \, 3e^{-2\mu}/3!, \, e^{-3\mu}/3!$,
so we expect that the frequency of occurrence in random
sampling should be $1:3e^{-\mu}:3e^{-2\mu}:e^{-3\mu}$.
As we see in table \ref{tab2},  
the results of Monte-Carlo simulations
are in perfect agreement with this expectation.

\begin{table}[h]
\begin{center}
\begin{tabular}{|c|c|c|c|c|c|}
\hline
& Graphs & \multicolumn{4}{|c|}{\includegraphics[width=5cm]{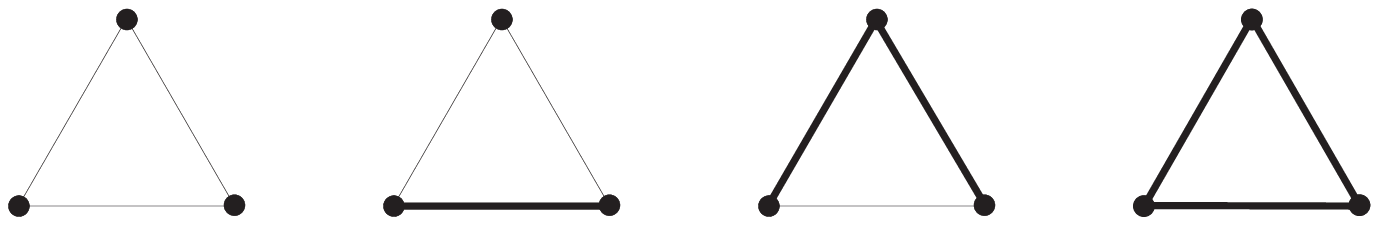}}	\\
\hline
$\mu = 0$ & Theor. & \ \ 0.125 \ \ & \ \ 0.375 \ \ & \ \ 0.375 \ \ & \ \ 0.125 \ \ \\
& Exp. & 0.125(1) & 0.374(1) & 0.374(1) & 0.126(1) \\
\hline
$\mu = 0.2$ & Theor. & 0.166 & 0.408 & 0.334 & 0.091 \\
& Exp. & 0.166(1) & 0.408(1) & 0.334(1) & 0.091(1) \\
\hline
$\mu = 0.5$ & Theor. & 0.241 & 0.439 & 0.266 & 0.054 \\
& Exp. & 0.241(1) & 0.439(1) & 0.266(1) & 0.054(1) \\
\hline
\end{tabular}
\caption{Comparison of the probability distribution
of graphs in a grand-canonical ensemble with $N=3$ 
nodes: calculated theoretically
and computed in a run of Monte-Carlo simulation in which
$10^6$ graphs were generated.}
\label{tab2}
\end{center}
\end{table}
One can easily apply this technique to any form of statistical
weights. In particular we can consider the product weights (\ref{eq:prod}).
The probability of accepting a new configuration by adding
or removing a link between $ij$ reads
\begin{eqnarray}
\mbox{min} \left\{ 1, \frac{N^2}{2(L+1)} 
\exp(-\mu) \frac{p(q_i+1)p(q_j+1)}{p(q_i)p(q_j)} \right\} & 
\mbox{for adding a link,} \nonumber \\
\mbox{min} \left\{ 1, \frac{2L}{N^2} 
\exp(+\mu) \frac{p(q_i-1)p(q_j-1)}{p(q_i)p(q_j)} \right\} &
\mbox{for deleting a link,} \nonumber
\end{eqnarray}
where $L$ and $q_i,q_j$ refer to the current configuration.

\section{Monte-Carlo generator of micro-canonical ensemble}
\begin{figure}[h]
\psfrag{i}{$i$} \psfrag{j}{$j$} \psfrag{k}{$k$} \psfrag{l}{$l$}
\includegraphics[width=5cm]{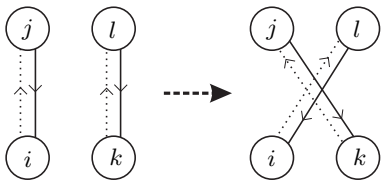}
\caption{The idea of ''X-move'': two oriented links (dotted lines) 
$ij$ and $kl$ chosen in a random way are rewired, exchanging their endpoints.
Then the opposite links (solid lines) are also rewired. }
\label{fig:xmove}
\end{figure}

Another frequently encountered ensemble is an ensemble of graphs
which have a given node degree sequence $\{q_1,q_2,\dots,q_N\}$.
The partition function $Z$ has the form:
\bq
Z(N,\left\{ q_i \right\}) = \sum_{\alpha'\in lg(N,L)} 
\left( \prod_{i=1}^N \delta \left[ q_i(\alpha') - q_i\right] \right) 
\, 1/N! \, W(\alpha') ,
\eq
where the product of delta functions allows one to include only 
those graphs which have a prescribed degree distribution $q_i$.
As before the factor $1/N!$ is fixed in this ensemble
and could in principle be skipped.
%One sees that $Z(N,L)$ for canonical ensemble can be easily 
%obtained from
The canonical partition function $Z(N,L)$ is related to
the micro-canonical ones:
\bq
Z(N,L) = \sum_{q_1=0}^\infty \dots \sum_{q_N=0}^\infty 
Z(N,\left\{ q_i \right\}) \, \delta \left[ q_1+q_2+\dots +q_N - 2L \right] .
\eq
To generate graphs from micro-canonical ensemble one has to
have a Markov process preserving node degrees. 
The main idea is to combine
simultaneous rewirings \cite{ref:msz}
as shown in Fig. \ref{fig:xmove}.
We shall call this combination ''X-move''.
At each step one picks up two random links: $ij$ and $kl$, 
and rewires them to $il$ and $kj$. 
This procedure is ergodic, i.e. it explores the whole 
configuration space. Such a transformation
was discussed in \cite{ref:msz} where it was used to 
randomize graphs with a given nodes' degree sequence. 
In that case the functional weight was $W_\alpha=1$ and
rewirings were done with probability equal to one. 
In general case if one considers non-trivial 
$W_\alpha$, one has to accept the change with 
a corresponding Metropolis probability (\ref{metrop}).
In this way one can for example generate graphs
whose statistical weights depend on the number 
of triangles. In a sense one can perform
a weighted randomization of networks with the given
node degree sequence. Introducing a weight
into randomization may be very important in the
construction of scoring functions in problems 
of motif searching \cite{ref:m,ref:bl2,ref:ia}. 
If one tries to determine relations between structural motifs 
and the functionality of network, it is very important 
to properly construct scoring function which may
clearly account for the existence of a particular
subgraph on a network and its function. Scoring functions
are usually measured as a sort of distance between 
a network which displays a certain function and 
a random network which does not. 
An important problem in such studies is how to construct those
networks which should serve as the background reference.
The simplest idea is to use networks obtained
by uniform randomization. This may however introduce some bias 
and may be misleading. Imagine
for example that a motif which is responsible for a certain network
function is built out of a couple of triangular loops and
that triangular loops alone have no function. 
It is clear that one would like to control the abundance 
of triangular loops to distinguish between specific motifs and
motifs which are more frequent by pure chance just because of higher
abundance of triangles. Therefore it may be important to control
the number of triangles in the randomized reference networks
used in the scoring function. It was just an example, but in
general case it might be useful to perform a weighted randomization
taking into account some desired features of reference networks.

\section{Graph generator and adjacency matrices}

All the elementary transformations: rewiring, adding or removing a link,
and the X-move have a simple representation in terms of adjacency
matrices. Rewiring relies on picking up at random an element 
$A_{ij} = 1$ of the adjacency matrix and flipping it with
an element $A_{ik}=0$ so that after the move 
$A_{ij} = 0$ and $A_{ik}=1$. For undirected links
adjacency matrices are symmetric and therefore at the same time 
one has to flip $A_{ji} = 1$ and $A_{ki}=0$.  
To add a vertex one chooses at random $A_{ij}$ and if
$A_{ij}=0$ and $i\ne j$, one changes it into $A_{ij}=1$ 
(and for $A_{ji}$ correspondingly).
To remove a link one picks up a non-vanishing element
$A_{ij}=1$ and substitutes it with $A_{ij}=0$. To perform
X-move one picks up two non-vanishing elements of $\A$ at random,
say $A_{ij}=1$ and $A_{kl}=1$, and flips them simultaneously with
$A_{il}=0$ and $A_{kj}=0$ to:
$A_{ij}=0$, $A_{kl}=0$, $A_{il}=1$ and $A_{kj}=1$.
Of course one also flips their four symmetric
counterparts. In practice, when simulating sparse graphs
one does not use the matrix
representation since it would require $N^2$ storage capacity.
For sparse matrices the number of non-vanishing
matrix elements is proportional to $N$ and one can
use a linear storage structure. It directly corresponds to the underlying
graph structure. Using linear storage one can code graphs having of
order $10^6$ nodes or even more on a PC.

\section{Degenerated graphs (pseudographs)}

In previous sections we described ensembles of simple
graphs. Let us now discuss pseudographs that is graphs
which may have multiple- and self-connections. 

\begin{figure}[h]
\psfrag{1}{$1$} \psfrag{2}{$2$} \psfrag{3}{$3$} \psfrag{4}{$4$} \psfrag{5}{$5$} \psfrag{6}{$6$}
\includegraphics[width=7cm]{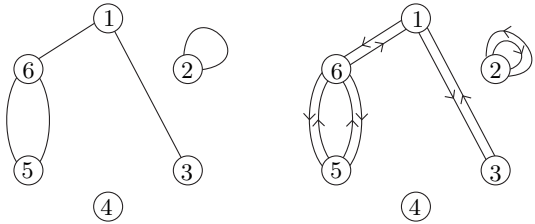}
\caption{Left hand side: the example of pseudograph with 
$N=6,L=5$. Right hand side: its representation as directed graph.}
\label{fig:pseudo}
\end{figure}

A degenerate undirected pseudograph can be represented 
by a symmetric adjacency matrix $\A$ whose off diagonal entries 
$A_{ij}$ count the number of links between vertices $i$ and $j$,
and the diagonal ones $A_{ii}$ count twice the number of
self-connecting links attached to vertex $i$.
For example, the graph depicted in Fig. \ref{fig:pseudo} 
has the following adjacency matrix:
\bq
\A = \left( \begin{array}{cccccc}
0 & 0 & 1 & 0 & 0 & 1 \\
0 & 2 & 0 & 0 & 0 & 0 \\
1 & 0 & 0 & 0 & 0 & 0 \\
0 & 0 & 0 & 0 & 0 & 0 \\
0 & 0 & 0 & 0 & 0 & 2 \\
1 & 0 & 0 & 0 & 2 & 0
\end{array} \right).
\label{eq:pseudo}
\eq
%The diagonal elements of the adjacency matrix
%count the number of self-connecting links attached to
%the corresponding vertex multiplied by the factor of two. 
In the representation where each undirected link is 
represented as two opposite directed links 
all matrix elements including the diagonal ones 
count the number of directed links emerging 
from the vertex.

As before let us first consider labeled pseudographs. 
However, in order to have a unique representation of a graph
one has to label links as well. We did not have to do
this for simple graphs since in that case each link was uniquely
determined by its endpoints. It is not anymore the case for
degenerate graphs since there may be more than one link 
between two nodes. A pseudograph
with $N$ nodes and $L$ links can be fully labeled
by $N$ node labels and $2L$ labels of directed links. 
Each fully labeled graph has
thus the configuration space weight equal to $1/(N!(2L)!)$. 
Let us work out the consequences of this choice. 
Denote $\alpha$ a graph, $\alpha'$ a labeled
graph of $\alpha$ with labeled nodes only, and $\alpha''$
a fully labeled graph of $\alpha$ with labeled nodes and 
labeled links. From here on labeled graph means a graph which has only 
labels on nodes while \emph{fully labeled graph} a graph which
has additionally labels on directed links.

The configuration space weight of $\alpha$ can be calculated
as a sum over all fully labeled graphs $\alpha''$ as follows:
\bq
w_\alpha = \sum_{\alpha'' \in flg(\alpha)} \frac{1}{N!(2L)!} =
\sum_{\alpha' \in lg(\alpha)} \frac{1}{N!} 
\left( \prod_{i} \frac{1}{2^{A_{ii}/2} \left( A_{ii}/2\right)!} \right)
\prod_{i>j} \frac{1}{A_{ij}!} ,
\label{wp}
\eq
where $flg(\alpha)$ denotes the set of fully 
labeled graphs of graph $\alpha$, $lg(\alpha)$ the set of 
labeled graphs (labeled nodes only) of graph $\alpha$ . 
The expression $A_{ii}/2$ counts the number of 
self-connecting links attached to vertex $i$,
and $A_{ij}$ is the multiplicity of links connecting $i$ and $j$.
If there are no self-connections ($A_{ii}=0$)  
and no multiple connections ($A_{ij}\le 1$), the configuration
space weight (\ref{wp}) reproduces the weight of simple graphs.
One can easily understand the appearance of the 
combinatorial factors in general case. Suppose that
we permute links' labels of a fully labeled graph 
leaving nodes' labels intact. Among all $(2L)!$ permutations
not all are distinct. If we have $A_{ij}$ links
between vertex $i$ and $j$ and we will permute their labels, then
all $A_{ij}!$ permutations will give the same labeled 
graph (if we simultaneously permute labels of the directed partners).
Similarly if we have vertex with a self-connection and 
we exchange labels of the two directed links emerging from
this vertex, the fully labeled graph will not change. Thus
for each self-connection two permutations lead to the same
fully labeled graph. To summarize, the number of distinct
permutations of link labels is reduced from $(2L)!$ by dividing
out the factor $2$ for each self-connection and $k!$ for each $k$-link
multiple connection which just gives Eq. (\ref{wp}).
It turns out that these weights
are identical to the combinatorial factors of 
Feynman diagrams which appear in perturbative
series of a mini-field
theory \cite{ref:bck}. One can thus interpret random
pseudographs as Feynman diagrams and use perturbation theory
to enumerate them. 

\begin{figure}[h]
\includegraphics[width=12cm]{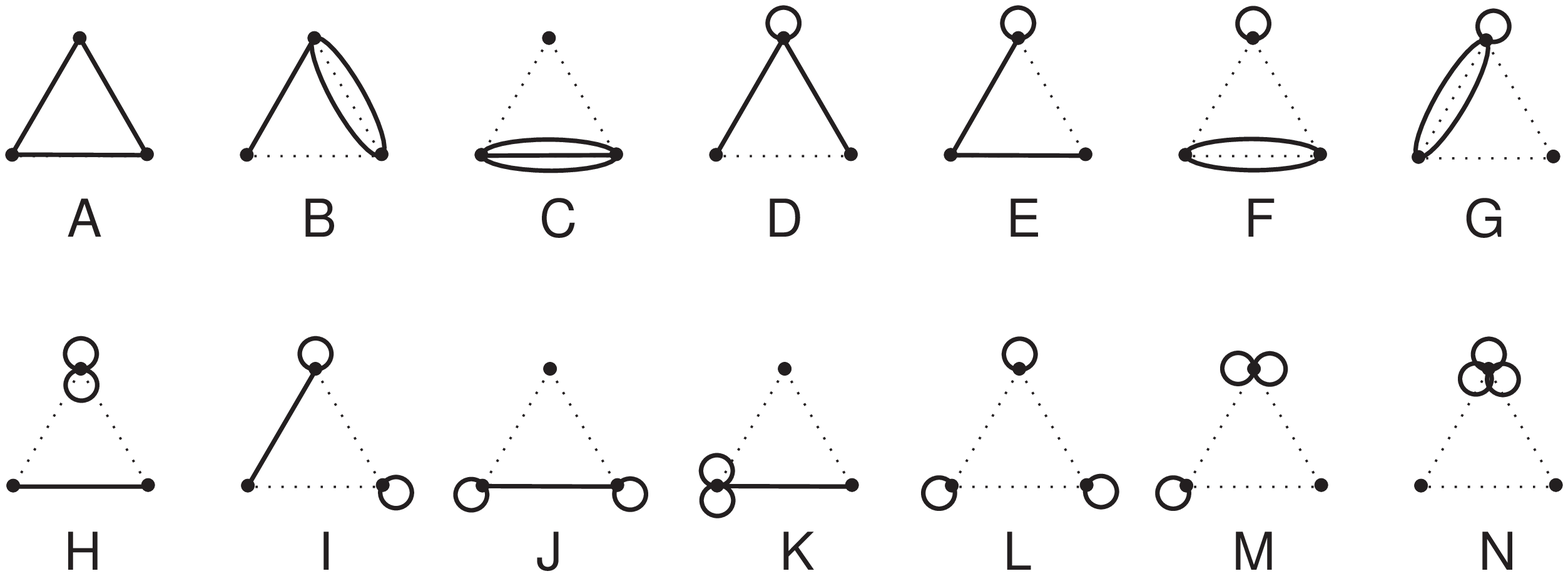}
\caption{All pseudographs in the canonical ensemble with $N=3,L=3$.}
\label{fig:pseudo_3}
\end{figure}

Let us consider as an example a canonical ensemble of pseudographs
with $N=3$ and $L=3$. There are $14$ graphs in this ensemble.
They are shown in Fig. \ref{fig:pseudo_3}.
In table \ref{tab3} we compare the 
theoretically calculated probability distribution
of graphs: 
\bq
p_\alpha = \frac{w_\alpha}{\sum_\beta w_\beta}  ,
\eq
using the weights calculated by the formula (\ref{wp}) 
with the probability distribution obtained experimentally
from the frequency histogram
of graphs produced by the Monte-Carlo generator.
Now the generator works exactly as before except that
it does not reject moves leading to a self- or multiple-connections.
The results are in perfect accordance.

\begin{table}[h]
\begin{center}
\begin{tabular}{|c|c|c|c|c|c|c|c|}
\hline
Graphs & A & B & C & D & E & F & G \\
\hline
Weights & 1/6  & 1/2 & 1/12 & 1/4 &  1/2 &  1/8 & 1/4 \\
\hline
Theor.  & 0.066  & 0.197 & 0.033 & 0.099 & 0.197  & 0.049  & 0.099 \\
\hline
Exp. & 0.066(1)  & 0.197(1) & 0.033(1) & 0.099(1) & 0.197(1) & 0.049(1) & 0.099(1) \\
\hline	
\hline
Graphs & H & I & J & K & L & M & N 	\\
\hline
Weights & 1/16 & 1/4 &  1/8 & 1/8 & 1/48 & 1/16 & 1/96 \\
\hline
Theor.  & 0.0247 & 0.099 & 0.049 & 0.049 & 0.0082 & 0.0247 & 0.0041 \\
\hline
Exp. & 0.0246(1) & 0.099(1) & 0.049(1) & 0.049(1) & 0.0082(1) & 0.0247(1) & 0.0041(1) \\
\hline		
\end{tabular}	
\caption{Comparison of theoretical and experimental (Monte-Carlo) computations of 
frequencies of graphs' occurrence in the ensemble with $N=3,L=3$ \label{tab3}.} 
\end{center}
\end{table}

As an example let us calculate the weight of graph M in Fig. \ref{fig:pseudo_3}. 
The weight of each labeled graph of graph M,
according to the formula (\ref{wp}), is equal
\bq
w_{M'} = \frac{1}{3!} \cdot \frac{1}{2^3} \frac{1}{2!} = \frac{1}{96}  ,
\eq
where the first factor comes from $1/N!$, the
second from the three self-connections, and the third
from the fact that the two self-connections are attached
to the same vertex and thus can be permuted without changing 
graph's connectivity. There are six distinct labeled graphs M'
of graph M and thus
\bq
w_{M} = \sum_{M'} \frac{1}{96} = \frac{6}{96} = \frac{1}{16}  .
\eq
One should note that the number of distinct labeled graphs
varies from graph to graph. For example, for graph L there is only
one labeled graph. In this case $w_{L'} = 1/3! \cdot 1/2^3 = 1/48$
and $w_L = 1/48$. The calculation can be easily repeated 
for each graph in Fig. \ref{fig:pseudo_3} 
yielding the weights $w_\alpha$ listed
in table \ref{tab3}. 

As follows from Eq. (\ref{wp}),
the partition function for the canonical ensemble of
pseudographs can be written in three different ways:
\bq
Z(N,L) = \sum_{\alpha'' \in flg(N,L)} \frac{1}{N! (2L)!} \ = 
\sum_{\alpha' \in lg(N,L)}  \frac{1}{N!}
\left( \prod_{i} \frac{1}{2^{A_{ii}/2} \left( A_{ii}/2\right)!} \right)
\prod_{i>j} \frac{1}{A_{ij}!} \  = 
\sum_{\alpha \in g(N,L)} w_\alpha  .
\label{zw}
\eq
The first sum runs over fully labeled graphs and has the
simplest form since all fully labeled graphs have the same
weight. The weight of labeled graphs in the second sum
varies. We note that labeled graphs are isomorphic with
adjacency matrices that is each labeled graph is
uniquely represented by a generalized adjacency 
matrix \footnote{The elements $A_{ij}$ are indexed by
nodes' labels, but the information about links' labels
is lost in the adjacency matrix representation.}
like Eq. (\ref{eq:pseudo}). The
sum over $\alpha' \in lg(N,L)$ can be thus interpreted
as a sum over all generalized adjacency
$N \times N$ symmetric matrices $\A$ 
such that $\sum_{ij} A_{ij} = 2L$. We see that 
not all adjacency matrices have the same statistical weight
unlike for simple graphs since the weights depend on
the number of self- and multiple-connections.

A weighted ensemble of pseudographs is constructed as before
by introducing an additional functional weight $W(\alpha)$
under the sum defining the partition function (\ref{zw}):
\bq
Z(N,L) = \sum_{\alpha'' \in flg(N,L)} \frac{1}{N! (2L)!} W(\alpha'') \ =
\sum_{\alpha \in g(N,L)} w_\alpha W(\alpha)  .
\eq
As before the functional weight $W(\alpha'')$ does not depend 
on graph's labeling but only on graph's topology. 
In other words
if $\alpha''_1$ and $\alpha''_2$ are two different fully
labeled graphs of graph $\alpha$ then
$W(\alpha''_1)=W(\alpha''_2) \equiv W(\alpha)$.

We can now consider various weights: for example a
product weight as in Eq. (\ref{eq:prod}) to mimic 
graphs with uncorrelated node degrees.
But even in this case the total weight does not factorize since
the configuration space weight $w(\alpha)$ written 
as a function of node degrees $w(q_1,q_2,\dots, q_N)$ 
does not factorize. Due to the absence of the structural
constraints the approximation given by equations (\ref{fact}) and 
(\ref{piqAB}) has now much weaker finite size corrections.

A grand-canonical ensemble for pseudographs 
with arbitrary product weights (\ref{eq:prod}) 
has the following partition function:
\bq
Z(N,\mu) = \sum_L \exp (-\mu L) 
\sum_{\alpha \in g(N,L)} w_\alpha  \prod_{i=1}^N p(q_i(\alpha))  .
\label{eq:grps}
\eq

This means that all pseudographs with fixed nodes' degrees $\{q_i\}$
have the same functional weight $\sim p(q_1)\cdots p(q_N)$,
which seems to be similar to
that generated by the Molloy-Reed
construction of pseudographs \cite{ref:mr}. 
Let us comment on this.
In the Molloy-Reed construction
one generates a sequence of non-negative integers $\{q_1,q_2,\dots,q_N\}$
for example as independent identically distributed
numbers with the distribution $p(q)$. One interprets $q_i$'s as node degrees. 
The only requirement is that the sum $q_1+q_2+\dots +q_N=2L$ is even.
In the first step of the construction
each integer $q_i$ is represented as a hub built out of
a vertex and $q_i$ outgoing branches which can be viewed 
as directed links emerging from this vertex. 
In the second step the directed links are paired randomly 
in couples of links in opposite direction to
form undirected links connecting vertices. 
%Thus for given $\{q_1,q_2,\dots,q_N\}$ 
This procedure generates
the same subset of pseudographs as 
the partition function $Z(N,\mu)$ (\ref{eq:grps}).
However, statistical weights are different.

To see this, let us consider a subset of Molloy-Reed graphs obtained for
a given set $\{q_i\}$.
There are $N$ labeled
vertices and $2L=\sum_i q_i$ labeled directed links. All permutations of 
labels of links and nodes are equiprobable   
exactly as it was before for fully labeled pseudographs (\ref{wp}). 
If one calculates corresponding symmetry factors
for node-labeled graphs the same combinatorial factors arise
as in Eq. (\ref{wp}): if one pairs two directed links $a$ and $b$
which belong to the same vertex one obtains a self-connecting link.
The pair $ab$ is identical as $ba$ 
since both the links begin and end at the same vertex.
This reduces the number of distinct permutations by factor of $2$
as in Eq. (\ref{wp}). Similarly for $k$ pairs of directed links between two 
vertices one can exchange the order of pairing in $k!$ ways
each time obtaining the same multiple connection, so the corresponding
factor is $1/k!$ again as in Eq. (\ref{wp}).
The conditional probability of choosing a particular graph 
$\alpha$ under the condition that in the first step of the construction 
the set of $\{q_1,q_2,\dots,q_N\}$ has been selected, is
\bq
w_{M-R}(\{q_i\}|\alpha) = 
\frac{w_\alpha}{\sum_{\beta \in g\{q_1,\dots ,q_N\}} w_{\beta}}  ,
\eq
where the sum is done over all (unlabeled) 
pseudographs $\beta$ from the micro-canonical set
of fixed degrees $\{q_1,\dots ,q_N\}$.
The total probability is thus
\bq
w_{M-R}(\alpha) = P(\{q_i\}) \, w_{M-R}(\{q_i\}|\alpha)  ,
\eq
where $P(\{q_i\})$ is the probability that
in the first step of the construction the set 
$\{q_1,\dots ,q_N\}$ is selected.
This probability is proportional to the product 
of $p(q_i)$'s multiplied by the number of permutations of
$\{q_1,\dots ,q_N\}$ which give the same set.
We denote this number by $\mbox{Perm}(q_1,\dots, q_N)$.
The order of $q_i$'s does not matter since we consider unlabeled 
graphs. For example, the following
permutations (sequences): $(q_1,q_2,q_3)=(3,3,2)$,
$(3,2,3)$ and $(2,3,3)$ give the same set $\{3,3,2\}$,
so in this case we have $\mbox{Perm}(3,3,2)=3$.
In general, the number is given by
\bq
\mbox{Perm}(q_1,\dots, q_N) = \frac{N!}{n_0! n_1!\cdots}  ,
\eq
where $n_0,n_1,\dots$ are degree's multiplicities:
$n_q=\sum_i \delta\left[ q_i-q\right]$. Thus
\bq
P(\{q_i\}) \propto
\left( \prod_{i=1}^N p(q_i(\alpha)) \right) \mbox{Perm}(\{q_i\}) .
\eq
\begin{figure}[h]
\includegraphics[width=12cm]{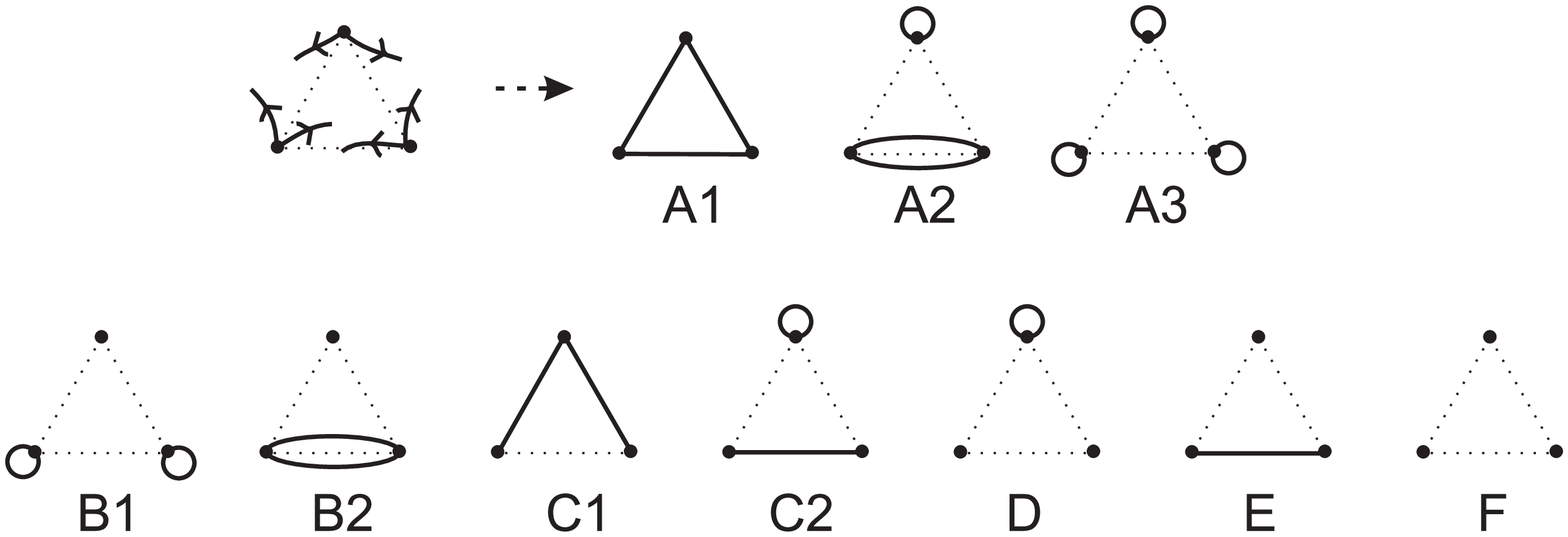}
\caption{A set of 10 pseudographs for $N=3$ and $p(q)=1/3$ for
$q=0,1,2$. Top: three hubs for $\{q_i\}=\{2,2,2\}$
are generated and then directed links are paired randomly giving three
pseudographs A1,A2,A3. Bottom: the rest of pseudographs from
this ensemble.}
\label{fig:grps}
\end{figure}
\begin{table}[h]
\renewcommand*{\arraystretch}{1.3}
\begin{center}
\begin{tabular}{|c|c|c|c|c||c|c|}
%\hline
%\multicolumn{7}{|c|}{$\sum_{k_1,k_2,k_3} 
%\mbox{Perm}(k_1,k_2, k_3) = 1+3+3+3+3+1 = 14$} \\
\hline
L & $\{q_i\}$ & $\mbox{Perm}(q_1,q_2,q_3)$ & Graphs & $w_\alpha$ weights (\ref{wp}) & $w_{M-R}$ weights (\ref{eq:wmr}) & G-C weights from Eq. (\ref{eq:grps}) \\
\hline \hline
%\rule{2mm}{0mm}
3 & 2,2,2 & 1 & A1,A2,A3 & $(\frac{1}{6}:\frac{1}{8}:\frac{1}{48}) = \frac{1}{15} (8:6:1)$ & 
$\frac{1}{14} (\frac{8}{15},\frac{6}{15},\frac{1}{15})$ & $e^{-3\mu}(\frac{1}{6},\frac{1}{8},\frac{1}{48})$ \\
\hline \hline
2 & 2,2,0 & 3 & B1,B2 & $\frac{1}{8}:\frac{1}{4}=\frac{1}{3}(1:2)$ &
$\frac{3}{14} (\frac{1}{3},\frac{2}{3})$ & $e^{-2\mu}(\frac{1}{8},\frac{1}{4})$ \\
\hline
& 2,1,1 & 3 & C1,C2 & $\frac{1}{2}:\frac{1}{4}=\frac{1}{3}(2:1)$ &
$\frac{3}{14} (\frac{2}{3},\frac{1}{3})$ & $e^{-2\mu}(\frac{1}{2},\frac{1}{4})$ \\
\hline \hline
1 & 2,0,0 & 3 & D & $\frac{1}{4}$ & $\frac{3}{14}$ & $e^{-\mu}\frac{1}{4}$ \\
\hline
& 1,1,0 & 3 & E & $\frac{1}{2}$ & $\frac{3}{14}$ & $e^{-\mu}\frac{1}{2}$ \\
\hline \hline
0 & 0,0,0 & 1 & F & $\frac{1}{6}$ & $\frac{1}{14}$ & $\frac{1}{6}$ \\
\hline
\end{tabular}
\caption{Weights calculated for the Molloy-Reed construction
and for the corresponding grand-canonical ensemble with $N=3$.
For each sequence $q_1,q_2,q_3$ we give the combinatorial 
number $\mbox{Perm}(q_1,q_2,q_3)$ of permutations leading to 
the same graph. Altogether, there are $14$ different sequences
of length three, with $q_i=0,1$ or $2$ as can be seen in the
third column of the table.}
\label{tab:grps}
\end{center}
\end{table}
Collecting all the factors together and normalizing 
to have the probabilistic interpretation we obtain
the following expression for
the total weight (probability) for Molloy-Reed's 
pseudographs:
\bq
w_{M-R}(\alpha) = \left( \prod_{i=1}^N p(q_i(\alpha)) \right)
\frac{\mbox{Perm}(q_1(\alpha),\dots, q_N(\alpha))}
{\sum_{k_1,\dots, k_N} \mbox{Perm}(k_1,\dots, k_N)} \,
\frac{w_\alpha}{\sum_{\beta \in g\{q_1,\dots ,q_N\}} w_{\beta}}  .
\label{eq:wmr}
\eq
The first factor comes from picking $N$ numbers $q_i$ at random, 
the second counts permutations
and the third includes the weight generated by pairing directed links.
As we see, despite many similarities the Molloy-Reed ensemble
and the grand-canonical lead to different weights. 
As an example, in Fig. \ref{fig:grps} we show an ensemble of 10 pseudographs
with $N=3$ generated by Molloy-Reed algorithm for $p(q)=1/3$ for
$q=0,1,2$ and zero elsewhere. We compare statistical weights
of the generated graphs with the corresponding ones in the
grand-canonical ensemble. As we can see in table \ref{tab:grps},
the weights are different in the two ensembles.

\section*{Summary}

We have discussed a statistical approach
to homogeneous random graphs. This framework is 
a natural extension of the Erd\"os-R\'enyi theory 
to the case of weighted graphs: one considers the
same set of graphs but with modified statistical weights.
The statistical weights of homogeneous graphs depend 
only on graphs' topology. In other words, if one assigns
some labels to its nodes, they will
have no physical meaning similarly as the numbers
of indistinguishable particles in quantum mechanics.
One can permute them and the graph and its statistical
weight will stay intact. The only information which matters
is the number (entropy) of distinct permutations of nodes' labels.
All permutations of node's labels
are equivalent, unlike for growing
networks where those permutations have to preserve the
causal order corresponding to the order of node's attachment to the graph. 
The statistical weight of a homogeneous graph is proportional 
to the number of all labeled graphs of this graph while 
of a growing network to the number of causally labeled graphs. 
This leads to a difference between homogeneous and growing networks.
For example, a typical homogeneous graph has a larger diameter than
the corresponding growing network with the same node degree
distribution (\ref{eq:exact_p}). Generally, geometrical properties of homogeneous
graphs are different from those of growing networks for which
correlations between the time of node's
attachment and its degree induce 
node-node correlations of a specific type \cite{ref:kr,ref:bbjk}.
Such correlations are absent for homogeneous graphs. 

Various functional properties of homogeneous networks can 
be modeled by an appropriate choice of functional weight.
One can easily produce networks with an assertive mixing, 
higher clustering or any desired property which can
reflect any real-data observation. 

Homogeneous networks can be simulated numerically.
We have also described a  Monte-Carlo algorithm to 
generate canonical, grand-canonical and micro-canonical
ensembles which performs a sort of weighted random walk 
(Markov chain) in the configuration space with a desired
stationary distribution. We advocated the 
importance of the possibility of generating random 
networks with desired statistical properties 
for advanced motif searching \cite{ref:m,ref:bl2,ref:ia}.

Many real networks have resulted from hybrid processes
of growth mixed with some thermalization. The framework discussed
in this paper can flexibly extrapolate between the two regimes.
It allows one to directly investigate the relation between
structural and functional properties of complex networks.

\bigskip
\noindent
{\bf Acknowledgments}

\medskip
\noindent
We thank Piotr Bialas, Jerzy Jurkiewicz and Andrzej Krzywicki
for stimulating discussions. This work was partially supported by 
the Polish State Committee for Scientific Research (KBN) grant
2P03B-08225 (2003-2006) and Marie Curie Host Fellowship
HPMD-CT-2001-00108 and by EU IST Center of Excellence "COPIRA".

\end{document}